**Enhanced Photomultiplication Effect by Synergistic Integration of Hole-Blocking Layers and Trap Engineering in PM-OPDs**

*Awais Sarwar[1,]*, Louis Conrad Winkler[1,2], Anncharlott Kusber[1], Fred Kretschmer[1], Karl Leo[1]*, Hans Kleemann[1], Johannes Benduhn[1,2,]**

A. Sarwar, L.C. Winkler, A. Kusber, F. Kretschmer, K. Leo, J. Benduhn

[1] Dresden Integrated Center for Applied Physics and Photonic Materials (IAPP) and Institute of Applied Physics, Technische Universität Dresden, Nöthnitzer Str. 61, 01187 Dresden, Germany

[2] Zentrum für Technologieentwicklung, Deutsches Zentrum für Astrophysik, Postplatz 1, 02826 Görlitz, Germany

E-mail: awais.sarwar@tu-dresden.de, karl.leo@tu-dresden.de, hans.kleemann1@tu-dresden.de johannes.benduhn@tu-dresden.de

## Abstract

Photomultiplication-type organic photodetectors (PM-OPDs) promise exceptional sensitivity for weak-light detection but typically suffer from a gain-bandwidth trade-off where high external quantum efficiency (*EQE*) incurs large dark current and slow response times. Here, we demonstrate a fully vacuum-deposited PM-OPD architecture that mitigates these limitations by integrating hole-blocking layers low-stoichiometry molecular trap engineering. We isolate discrete trapping sites that maximize positive space-charge accumulation by introducing m-MTDATA as a dedicated hole-trapping site at a low concentration (0.5 wt%) into a BDP-OMe:$C_{60}$ bulk heterojunction. This engineered charge confinement triggers efficient field-assisted electron injection from the anode while remaining strictly below the threshold for localized percolation, effectively decoupling the photocurrent multiplication mechanism from trap-mediated dark current shunts. Consequently, the optimized device achieves a peak *EQE* exceeding 1100% at a reverse bias of -4 V. The optimized device exhibits a specific detectivity of $4\times10^{12}$ Jones under -2 V reverse bias along with a cutoff frequency ($f_{-3dB}$) of ~22 kHz.

## 1 Introduction

With the rapid advances in flexible electronics and photonics, organic photodetectors (OPDs) have become promising components for a variety of sensor applications. Compared to conventional inorganic photodetectors, OPDs offer several important advantages, including mechanical flexibility, low-cost fabrication, lightweight nature, and adjustable spectral response.[1,2] The applications that are of particular interest to these properties include the next generation uses in biomedical sensing, wearable electronics, high-resolution imaging, optical surveillance, and communications systems with growing demands for highly conformable sensor systems.[3] In particular, the OPDs have become increasingly targeted for applications in which detecting weak light signals is essential.[4,5]

Most conventional OPDs operate in the photovoltaic (PV) configuration, where they are only capable of reaching the physical limit of one collected electron per absorbed photon, which results in the limitation of the external quantum efficiency (*EQE*) at less than 100%. However, increasing responsivity (*R*), which is directly linked to *EQE*, can help to simplify the system design and to lower overall cost. Photomultiplication-type OPDs (PM-OPDs) are especially of interest in this regard since they enhance the low photocurrent internally, without additional external circuit elements. Since the first work by Hiramoto *et al*. in 1994[6] introducing OPDs with intrinsic gain, PM-OPDs have demonstrated signal amplification for several orders of magnitude.[7,8] In contrast to inorganic avalanche photodiodes, where gain arises from impact ionization, photomultiplication in OPDs generally relies on an imbalance between carrier lifetime and transit time.[9,10] This imbalance can originate from asymmetric carrier mobilities, trapping, or charge accumulation, analogous to photoconductive gain.[11,12] In PM-OPDs, this condition is commonly exploited by accumulating one carrier type near a contact or blocking layer, which modifies the local electric field and interfacial energy landscape.[13–15] Under reverse bias, this charge accumulation can promote field-assisted injection or generation of additional mobile carriers from the external circuit, leading to *EQE* values exceeding 100% when the accumulated-charge lifetime exceeds the transit time of the injected carrier.[16,17]

Although the PM-OPDs have shown remarkable performance improvements, several issues are yet to be addressed before they can be used in real-life, high-performance applications. While intentional charge traps are necessary for the photomultiplication effect and high responsivity, a high concentration of such traps may create continuous percolation pathways through the active material.[18,19] As a result, the charge carriers can undergo thermally activated hopping between neighboring trap sites, and in combination with trap-assisted charge generation,[20] form leakage channels that lead to a significant increase in the dark current density ($J_d$).[18,21,22] As a result, PM-OPDs that achieve high *EQE*s often exhibit reduced specific detectivity ($D^*$) due to elevated shot noise, representing a key performance trade-off. To address this bottleneck, the primary challenge is to decouple the interfacial fields required for photomultiplication from the leakage pathways that induce dark current. Furthermore, the dependence on charge accumulation and slow de-trapping dynamics also provides a fundamental speed limitation where high gain will often be accompanied by a slow response time (typically in the millisecond range to second range),[23] providing bandwidth in the 100 Hz range, thereby limiting their use in high-speed applications such as video imaging and optical communication.[19,24,25] Previous studies have carefully considered different methods to reduce such performance trade-offs in PM-OPDs.[7,26] *Trap engineering* of the photoactive layer has been one such strategy. This is usually done by blending an extremely disproportionate (e.g., 100:1 or 1:100) donor-acceptor (D:A) ratio to provide isolated domains of the minority component, which then serve as local traps for charges within the continuous percolation path of the majority component.[27] In the most widely studied PM-OPD architecture, poly(3-hexylthiophene) (P3HT) serves as the donor host matrix, with a small concentration (typically ~1 wt%) of fullerene derivatives such as $PC_{71}BM$ forming isolated electron-trapping domains. The large offset between the LUMO levels of P3HT and $PC_{71}BM$ enables efficient electron trapping, while the absence of continuous acceptor percolation pathways suppresses the dark current.[28–30] In addition to disproportionate D:A ratio strategies, various types of trap-forming additives, including organic small molecules, inorganic nanoparticles (NPs), and quantum dots (QDs), have been purposely introduced into the active layer as controlled charge-trapping centers.[31–35]

Besides the inherent trap manipulation, a significant amount of effort has been invested in controlling carrier injection and extraction at the electrode contacts through *interface engineering*, particularly by using carrier-blocking layers.[36,37] The introduction of interfacial blocking layers serves a dual function: under dark conditions, these layers present a high energy barrier that suppresses carrier injection and thus minimizes $J_d$. Under illumination, they promote the accumulation of photogenerated minority carriers at the interface, thereby inducing energy-level bending that facilitates field-assisted injection of the opposite carrier.[14] Recent studies have shown that hole-blocking layer (HBL) materials with deep HOMO levels can enhance hole accumulation and electron tunneling injection, yielding significant improvements in *EQE*.[16,37] Furthermore, precise control of the interfacial layer thickness, down to the atomic scale in the case of materials such as $Al_2O_3$, is critical for suppressing dark current leakage while preserving efficient field-assisted carrier injection under illumination.[38] The use of tailored interfacial layers, such as thin metal oxides (e.g., $MoO_3$) or organic interlayers (e.g., rubrene), can help balance the trade-off between gain and dark current, thereby improving $D^*$.[14,39,40]

Although there have been individual improvements in noise reduction and gain, little has been done to optimize them in combination.[41,42] Prior investigations have typically focused on optimizing either the bulk trap concentration or the interfacial blocking layer independently. However, the systematic use of a dedicated hole-trapping molecular additive at low stoichiometry within bulk heterojunctions, combined with deep-HOMO HBLs, to simultaneously amplify gain and suppress dark current has not been systematically investigated. The issue is not only to introduce traps or barriers, but to synchronize the respective functions: the bulk material should be able to provide enough photocarriers and energetically favorable hole-accumulation sites (*trap engineering*). At the same time, the interface should also be guaranteed to retain its rectifying function in the dark while promoting field-assisted carrier injection under illumination (*interface engineering*).

Here, we report a fully physical-vapor-deposited device architecture that integrates interfacial hole-blocking with controlled hole-trap formation in the BDP-OMe:$C_{60}$ active layer. $HAT(CN)_6$ and $C_{60}$ function as dual hole-blocking layers that suppress undesired charge injection under dark while promoting interfacial hole accumulation under illumination. Unlike its conventional use as a hole-transport material,[43,44] m-MTDATA is incorporated directly into the active layer as a hole-trapping molecular additive. Since its HOMO level (-4.9 eV) is 0.4 eV shallower than that of BDP-OMe (-5.3 eV) donor, its incorporation favors the formation of localized hole-trapping sites that promote charge accumulation without creating continuous transport pathways. Systematic variation of the m-MTDATA concentration show that the extent and distribution of these traps strongly affect the PM response. At -2 V, the optimized device achieves an *EQE* of 140% and a $D^*$ of $4\times10^{12}$ Jones, nearly one order of magnitude higher $D^*$ than the HBL-only reference. At −4 V, introducing 0.5 wt% m-MTDATA increases the *EQE* from ~350% for the HBL-only device to >1100%, indicating that the hole-trapping molecular additive strongly enhances the baseline PM response. Beyond the device physics, the exclusive use of physical vapor deposition (PVD) is a practical advantage, because it enables precise control over low trap concentrations and multilayer interfaces while remaining fully compatible with established vacuum-processing workflows for scalable manufacturing and industrial upscaling. These results show that bulk trap engineering and interfacial blocking can be integrated within a single, vacuum-processable device architecture to enhance gain while suppressing noise in PM-OPDs.

## 2 Results and Discussion

### 2.1 Material System and Stack Architecture

The active layer of the baseline OPD is made up of the donor BDP-OMe (boron dipyrromethene with CF3 on the meso-C) and the acceptor $C_{60}$ in a D:A weight ratio of 3:7. Figure S1 (Supporting Information) provides the respective chemical structures of the materials used in the device stack. To modulate charge retention and induce PM gain, m-MTDATA (4,4',4''-tris(3-methylphenylphenylamino)-triphenylamine) is introduced as a hole-trapping molecular additive into the active layer. The device architecture and absorption spectra of the active layer are shown in Figure 1a and Figure 1b, respectively. The respective energy level diagram presented in Figure 1c indicates that the HOMO level of m-MTDATA (-4.9 eV) is 0.4 eV above that of the BDP-OMe donor ( -5.3 eV). This energetic offset provides energetically favorable hole-retention sites relative to the BDP-OMe host, which enhance the capture and accumulation of photogenerated holes while having minimal impact on electron transport through the $C_{60}$ acceptor. To spatially confine these accumulated charges, deep HOMO level $HAT(CN)_6$ and $C_{60}$ were chosen as HBLs. The deep HOMO levels of $HAT(CN)_6$ (-9.5 eV) and $C_{60}$ (-6.4 eV) effectively block hole extraction toward the ITO electrode under reverse bias, promoting the accumulation of photogenerated holes, as demonstrated in our previous work.[37]

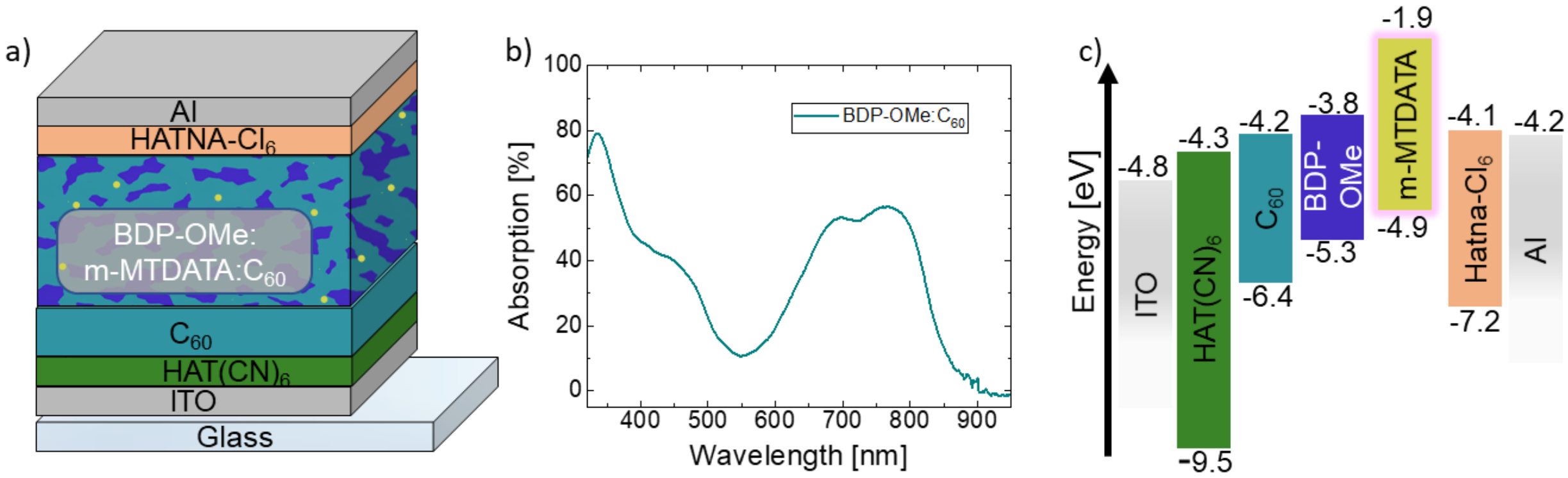

**Figure 1. Device architecture.** a) The optimized device architecture of the PM-OPD. b) Absorption of photoactive layer: BDP-OMe:$C_{60}$ (30 wt%, 100 nm). c) Energy level diagram for the device under dark conditions. The HOMO and LUMO values of the organic materials, HBLs, as well as the electrode's work function, are taken from the literature.[27,45,46]

Under reverse-bias conditions, this energy landscape supports the proposed trap-assisted PM mechanism as follows. The accumulation of holes induces significant interfacial energy-level bending, narrowing the injection barrier width at the ITO/HBL interface and enhancing electron injection from the external circuit. The $HAT(CN)_6$ layer was deposited with a thickness of 10 nm to balance efficient charge accumulation with field-assisted injection. Although hole accumulation governs the relatively fast turn-on dynamics of the PM-OPD, the overall temporal response is limited by the turn-off process, which depends on the much slower de-trapping and recombination rates of accumulated holes after illumination ceases.[14,37] As shown in Figure 1a, the device structure is ITO/$HAT(CN)_6$/$C_{60}$/BDP-OMe:m-MTDATA:$C_{60}$/HATNA-$Cl_6$/Al, where HATNA-$Cl_6$ is used as the electron transport layer (ETL). The entire device stack is fabricated by PVD, with an active area of 6.44 $mm^2$ defined by the overlap of the top and bottom electrodes. The specific details of the fabrication procedures are provided in the Experimental Section. The detailed influence of this trap-engineered architecture on the optoelectronic characteristics is analyzed in the following section.

**2.2 Role of trap states and hole-blocking layers in enabling photomultiplication**

To assess whether interfacial hole blocking is required for photomultiplication in this architecture, a series of control devices without HBLs were fabricated. In these control devices, m-MTDATA was incorporated at different concentrations (0.1 to 1.0 wt%). None of the control devices without the HBL exhibited *EQE* values exceeding 100%, even at -4 V reverse bias (Figure 1d), indicating the absence of PM effect. This observation indicates that under reverse bias, the presence of m-MTDATA-induced trap states alone is insufficient to produce PM gain in this device architecture; the interfacial hole-blocking layers are a prerequisite.

By preventing the extraction of photogenerated holes toward the ITO electrode, the HBLs facilitate the build-up of a positive space-charge region at the HBL/active layer interface.[14,47] This interfacial accumulation is expected to modify the local electric field and interfacial energy landscape, thereby enabling field-assisted electron injection from the external circuit under reverse bias. Without the blocking layers, photogenerated holes are likely extracted at the anode before sufficient space-charge accumulation and energy-level bending can develop, thereby preventing the onset of photomultiplication.

The operational mechanism under illumination and reverse-bias conditions is schematically illustrated in Figure 2a. Upon light absorption, excitons generated in the bulk heterojunction dissociate into free carriers. Photogenerated holes are preferentially retained at m-MTDATA-derived hole-accumulation sites because the m-MTDATA HOMO lies above that of BDP-OMe, while the deep HOMO levels of the HBLs suppress hole extraction toward the ITO electrode. As a result, this accumulation of positive space charge near the anode induces a sharp energy-level bending in the $HAT(CN)_6$ and $C_{60}$ layers. As illustrated in the diagram, this bending reduces the effective injection barrier at the ITO/HBL interface, promoting field-assisted electron injection from the ITO electrode into the device. These injected electrons are transported through the active layer to the Al cathode, producing a photocurrent that greatly exceeds the primary photogenerated current, thus yielding *EQE* > 100%.

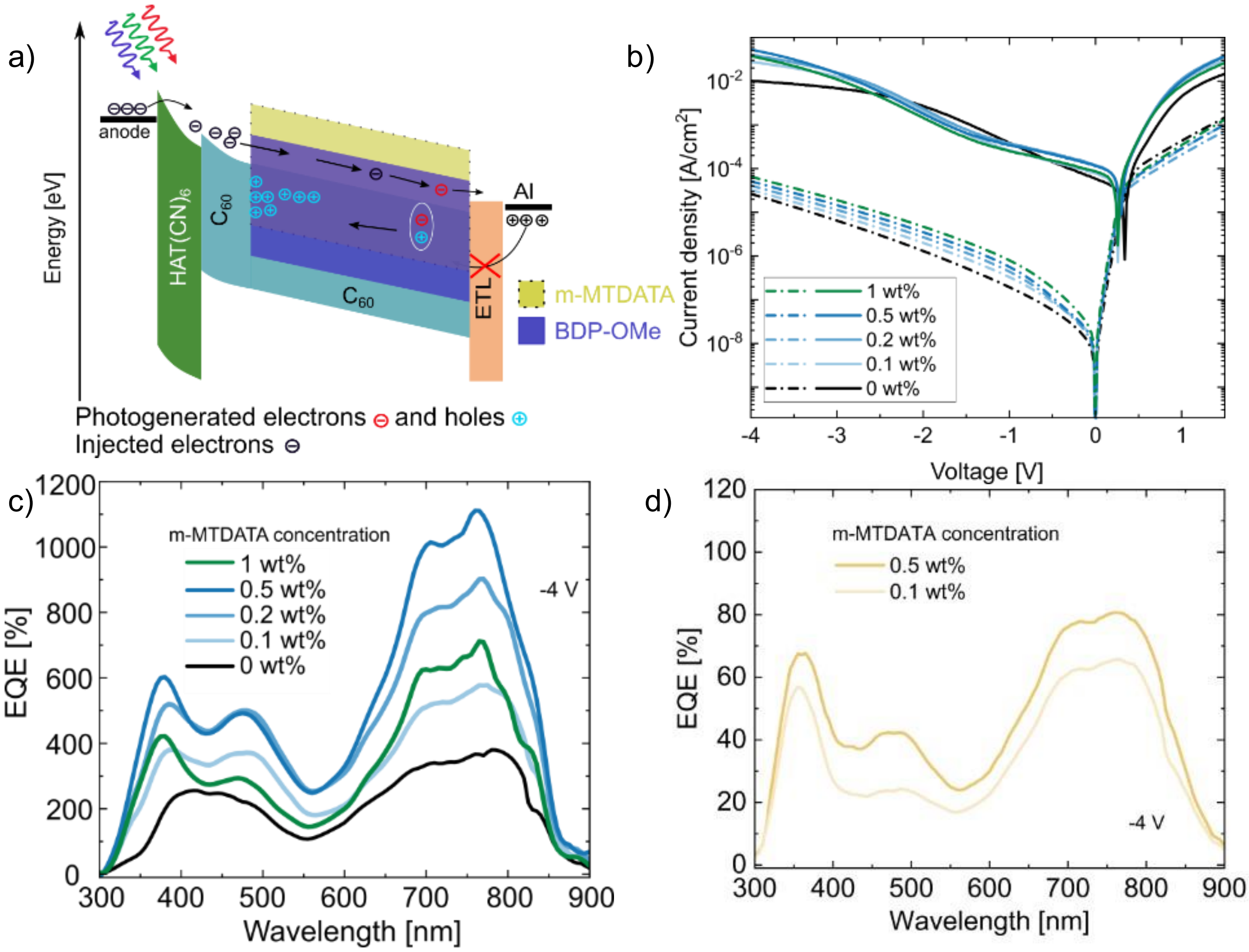

**Figure 2. Electrical characteristics of optimized PM-OPDs.** a) Operational mechanism of PM-OPD under illumination and reverse bias conditions. b) Current density vs. voltage curves in dark (dashed line) and under illumination (solid line) with a light intensity of 100 mW $cm^{-2}$. c) *EQE* spectra of PM-OPD with different concentrations of m-MTDATA. d) *EQE* spectra of devices with HBL under -4 V reverse bias.

To experimentally validate the trap-assisted gain mechanism and determine the optimal trap concentration, the current density-voltage (*J-V*) characteristics of the OPDs under both dark and illuminated conditions, are shown in Figure 2b for devices with varying m-MTDATA concentrations. In the dark (Figure 2b), the reference device (0 wt% m-MTDATA) exhibits a dark current of approximately $2\times10^{-5}$ A $cm^{-2}$ at -4 V, indicating that the $HAT(CN)_6$ and $C_{60}$ blocking layers effectively suppress charge injection from the electrodes. Upon incorporation of m-MTDATA, an increase in $J_d$ is observed as a monotonic function of increasing m-MTDATA concentration. The device with 1 wt% additive exhibits the highest dark current, reaching ≈ $7\times10^{-5}$ A $cm^{-2}$ at -4 V. This trend is consistent with the expected behavior of diluted trap systems: although the shallow HOMO of m-MTDATA provides energetically favorable hole-trapping sites, a high density of such sites reduces the inter-trap distance and can establish charge percolation pathways. These pathways can facilitate thermally activated hopping conduction in the dark and thereby moderately deteriorating the rectification behavior.[7,25]

Under illumination (Figure 2b), the device's photoresponse shows a pronounced dependence on m-MTDATA concentration. The m-MTDATA-containing devices exhibit a strong, non-saturating increase in photocurrent under reverse bias, consistent with photomultiplication. The photocurrent density increases substantially as the m-MTDATA concentration increases from 0.1 wt% to 0.5 wt%. However, further increase to 1.0 wt% leads to a decrease in the photocurrent density. Intentional m-MTDATA-induced trap states are therefore not a prerequisite for PM in this architecture, but rather amplify the HBL-enabled gain mechanism. The HBL-only reference already exhibits *EQE* values above 100%, while the optimized 0.5 wt% device shows stronger photocurrent amplification. The nonlinear onset of photocurrent under reverse bias is consistent with an energetic contact barrier whose partial screening under illumination supports the onset of PM. At excessive m-MTDATA concentration, recombination or percolative leakage can dissipate the accumulated space charge and reduce the net gain.

The existence of the PM effect and the optimization of the m-MTDATA concentration are further supported by the *EQE* spectra shown in Figure 2c (measured at -4 V bias). For the reference device containing HBL architecture but no m-MTDATA, the *EQE* reaches approximately ~350% across the measured spectral range, demonstrating that the HBLs alone already enable substantial PM gain. The introduction of m-MTDATA further amplifies this response, with the maximum *EQE* exceeding 1100% at approximately 780 nm for the 0.5 wt% device. This indicates that 0.5 wt% m-MTDATA provides the most favorable trap concentration in the investigated series.

At the maximum concentration of 1.0 wt%, the *EQE* decreases markedly and falls below that of the 0.1 wt% device. This non-monotonic trend highlights a critical trade-off in trap engineering. Up to 0.5 wt%, increasing the m-MTDATA concentration enhances positive space-charge build-up near the interface, strengthening the local field and energy-level bending that support photomultiplication. At 1.0 wt%, excessive m-MTDATA concentration likely promotes non-radiative recombination between accumulated holes and free electrons and may create percolative leakage pathways. These processes can reduce the effective positive space-charge density near the HBL/active-layer interface, and thereby suppress photomultiplication.[16]

**2.3 Performance metrics of the PM-OPDs**

Based on the concentration-dependent analysis above, the device containing 0.5 wt% m-MTDATA was selected as the optimized PM-OPD. To evaluate its practical suitability for weak-light detection, the noise spectral density, specific detectivity, and frequency response were compared with those of the reference device without m-MTDATA.

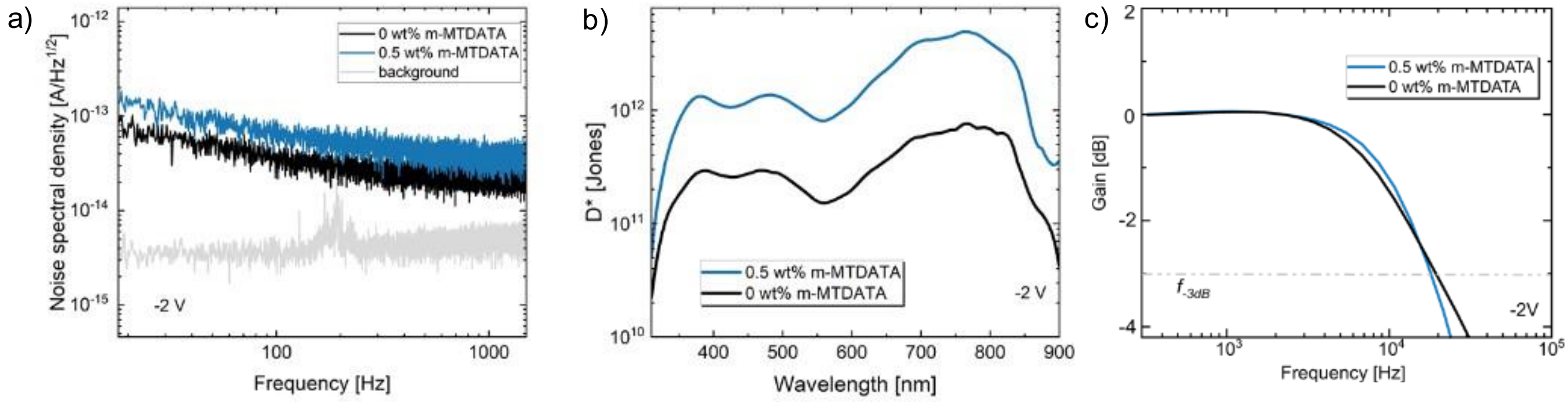

**Figure 3. Figures of merit.** a) Noise spectral density of OPDs with and without m-MTDATA under -2 V reverse bias, and the background measurement of the sample holder without a mounted sample. b) Specific detectivity of OPDs with and without m-MTDATA under -2 V reverse bias; based on noise measurements from a). c) Frequency response of the reference device without m-MTDATA and the device with 0.5 wt% of m-MTDATA at an illumination wavelength of 660 nm with an illumination intensity of 31.8 mW $cm^{-2}$.

High-gain photodetectors commonly face a trade-off: the introduction of trap-related states leads to an increase in dark current and noise that can offset the benefits of enhanced responsivity.[48] In the present devices, this trade-off is favorably managed at -2 V reverse bias. As shown in Figure 3a, a slight increase in the measured noise current density from $3.4 \times 10^{-14}\ A\,Hz^{-1/2}$ in the control device to $4.9 \times 10^{-14}\ A\,Hz^{-1/2}$ in the 0.5 wt% m-MTDATA device was observed at -2V. This modest increase is consistent with additional trap-related trapping and de-trapping processes introduced by m-MTDATA.[49]

In contrast to the small penalty in noise, the photo-response shows a pronounced enhancement. At -2 V, the *EQE* increases from 22% (reference device w/o m-MTDATA) to 140% (0.5 wt% m-MTDATA), as shown in Figure S2, indicating that the PM mechanism is operative at modest reverse bias. The specific detectivity $D^*$ was calculated based on the measured noise and responsivity using the expression:

$$D^* = \frac{R \cdot \sqrt{A}}{i_n} \tag{1}$$

where $R$ is the responsivity, $A$ is the active area, and $i_n$ is the measured noise current.[50] Consequently, the 0.5 wt% device achieves a peak $D^*$ of $4\times10^{12}$ Jones as shown in Figure 3b, representing nearly an order-of-magnitude improvement over the reference device with $D^*$ of $6\times10^{11}$ Jones. This result demonstrates that m-MTDATA enhances the photocurrent signal without a proportional increase in noise at -2 V, thereby improving the detectivity of the device. For comparison, the shot-noise-limited specific detectivity of the optimized device was also estimated as a function of reverse bias and remains in the $10^{12}$ Jones range across the investigated voltage range (Figure S3).

The temporal response of the devices was characterized from the frequency-dependent photocurrent response at -2 V (Figure 3c). Both the HBL-only reference and the 0.5 wt% m-MTDATA device exhibit a flat frequency response up to approximately 1–2 kHz, followed by a roll-off. The $f_{-3dB}$ cutoff frequency is approximately 25 kHz for the reference device and 22 kHz for the optimized m-MTDATA device, indicating that the introduction of m-MTDATA slightly reduces the device bandwidth. This modest reduction is consistent with the expected gain–bandwidth trade-off in PM-OPDs, as the extended carrier lifetime associated with trap-assisted charge accumulation inherently limits the temporal response.[25,51] Nevertheless, the bandwidth achieved at -2 V compares favorably with many reported PM-OPDs operating at substantially higher reverse biases, and remains sufficient for applications including photoplethysmography and imaging.[52,53]

### 2.4 Correlation between trap distribution, charge-transfer states, and photomultiplication gain

To obtain evidence for sub-gap states associated with m-MTDATA incorporation, ultra-sensitive external quantum efficiency (*us-EQE*) measurements were performed at a reverse bias of -2 V (Figure 4a). To minimize differences arising from the internal quantum efficiency, all of *us-EQE* spectra were normalized to the main absorption peak at 1.6 eV, which corresponds to the BDP-OMe:$C_{60}$ absorption band, which is constant throughout the experiment. The reference device without m-MTDATA exhibits a steep decrease in normalized *us-EQE* below 1.5 eV, reaching values of approximately $10^{-7}$ at 0.9 eV. In contrast, upon incorporation of m-MTDATA, a pronounced sub-gap feature emerges in the energy range of 0.8–1.2 eV , with *us-EQE* values that are 2–3 orders of magnitude higher than the reference (0 wt%) device. Since this feature is absent in the 0.0 wt% device and increases systematically with m-MTDATA concentration, it is assigned primarily to charge-transfer (CT) state absorption involving m-MTDATA:$C_{60}$ interfaces (see energetics in Figure 1c).

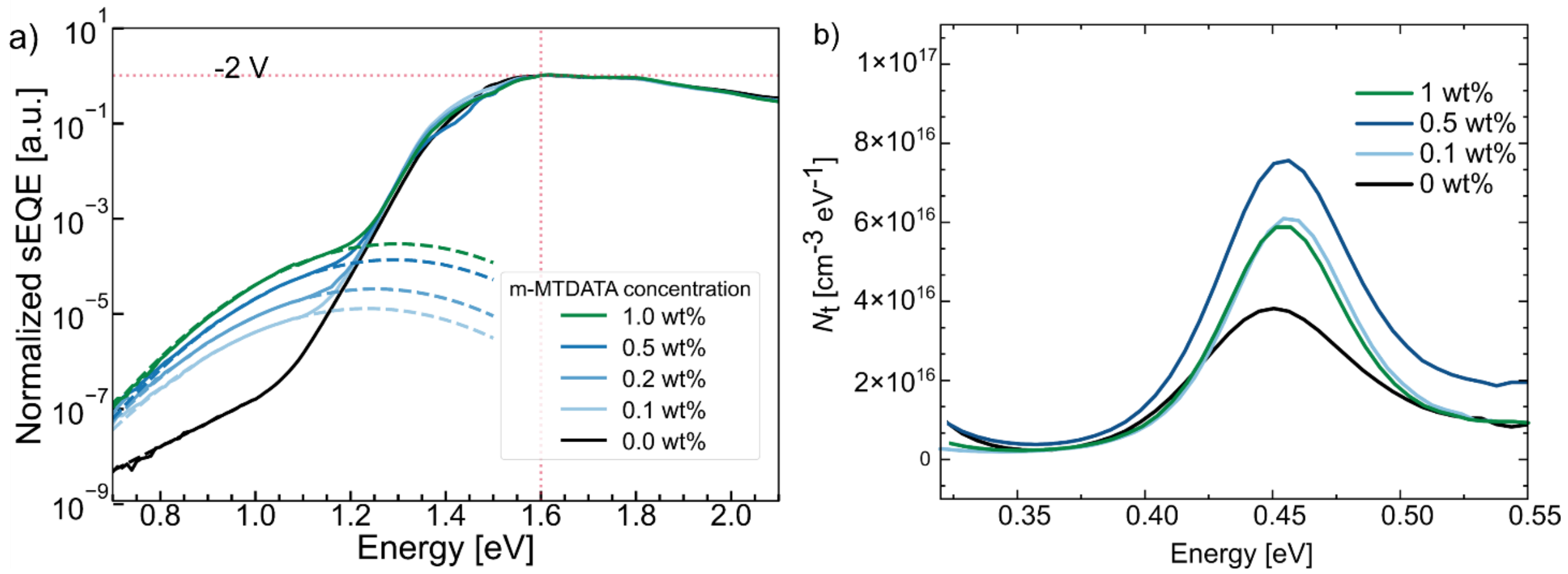

**Figure 4. Electrical and optical characterization of m-MTDATA-induced trap states.** (a) *Ultra-sensitive EQE* spectra at -2 V reverse bias. A distinct charge-transfer state absorption emerges between 0.8 and 1.2 eV upon m-MTDATA incorporation. (b) admittance-derived immobilized-charge/trap-related profiles for devices with varying m-MTDATA concentrations. Further details can be found in Figure S5.

To analyze this low-energy feature more quantitatively, the *us-EQE* response is fitted using a standard CT-state Marcus line-shape model.[54,55] The corresponding fit parameters are provided in Table S2 in the Supporting Information. The fitted CT-state energies, $E_{\mathrm{CT}}$, range from 0.78 to 0.88 eV. This spread is reasonable for disordered donor–acceptor blends, where variations in molecular conformation, packing, and interfacial microstructure give rise to a distribution of CT-state energies.[56] The reorganization-energy parameter, $\lambda_{\mathrm{CT}}$, remains between 0.44 and 0.47 eV. Both $E_{\mathrm{CT}}$ and $\lambda_{\mathrm{CT}}$ are in agreement with previous reports in the literature.[57] In addition, the fitted effective oscillator strength increases strongly with increasing m-MTDATA concentration, and scales approximately linearly with nominal m-MTDATA concentration, known from PVD process parameters. Since the control of the evaporation rate at very low concentrations (< 0.5 wt%) is technically challenging, the *us-EQE* data were used to calibrate the m-MTDATA concentration in the active layer; more details are provided in Figure S4.

The identification of m-MTDATA:$C_{60}$ CT absorption supports the proposed gain mechanism. Under reverse bias, dissociation of these CT states can generate holes on m-MTDATA sites and electrons $C_{60}$ network. Since the HOMO level of m-MTDATA lies approximately 0.4 eV above that of BDP-OMe, holes are energetically retained on m-MTDATA relative to the surrounding donor matrix, consistent with selective hole trapping and positive space-charge accumulation near the anode[19,40]. The concentration-dependent increase in CT absorption also follows the increase in dark current density, suggesting that these low-energy CT states contribute to field-assisted thermal charge generation in the dark.[1,58]

To further probe electrically active charge responses in the devices, temperature-dependent admittance spectroscopy was performed. This analysis identifies temperature- and frequency-dependent capacitance responses that are commonly associated with electrically active trap-related states. However, the method alone does not uniquely determine whether the response originates from trap-related states, interfacial capacitance, immobilized charges, or other slow charge-redistribution processes.[21,23,59,60] Figure 4b shows the admittance-derived immobilized-charge/trap-related profiles for devices with varying m-MTDATA concentrations.

The reference device without m-MTDATA already exhibits a weak feature near 0.45 eV, indicating that an electrically active charge response is present even in the BDP-OMe:$C_{60}$ device. Upon incorporation of m-MTDATA, the magnitude of the admittance-derived response increases and reaches its maximum at 0.5 wt%. The energetic position of this response is close to the nominal HOMO offset between m-MTDATA and BDP-OMe shown in Figure 1c, which is consistent with an additional contribution from m-MTDATA-related hole-retention sites. The temperature-dependent capacitance–frequency characteristics used for the admittance analysis are shown in Figure S5. Control devices without HBLs show only broad and weakly structured admittance-derived responses (Figure S6), suggesting that the HBLs enhance charge confinement and make the immobilized-charge/trap-related response more pronounced as we observe for devices with HBL and 0 wt% m-MTDATA.

The integrated admittance-derived response increases from the reference device to the 0.5 wt% m-MTDATA device, matching the concentration at which the strongest PM effect is observed. This trend suggests that optimized m-MTDATA concentration increases the electrically active immobilized-charge/trap-related response that can contribute to charge accumulation near the anode-side interface. At 1.0 wt%, however, the extracted electrically active discrete response decreases, even though the us-EQE data show that m-MTDATA-related CT states remain optically present. This indicates that higher m-MTDATA loading does not remove the CT states, but may reduce the fraction of states that behave as electrically isolated trap-like or immobilized-charged states within the admittance measurement window. As a result, the 1.0 wt% device exhibits stronger CT absorption while showing a weaker electrically active discrete response and reduced PM gain.

This analysis provides a plausible explanation for why the PM effect is strongest at 0.5 wt%: at this concentration, the m-MTDATA-induced states are sufficiently abundant to enhance positive space-charge accumulation, while still contributing effectively to the electrically active immobilized-charge/trap-related response. At 1.0 wt%, the CT states persist, but fewer of them contribute to this electrically active discrete response, which is consistent with the reduced PM efficiency. The reduced trapping efficiency at 1.0 wt% is connected due to the onset of hole percolation at this concentration which is consistent with similar reports in the literature.[18,61,62]

## 3. Conclusion

We demonstrate a PM-OPD architecture that amplifies PM gain by engineering discrete hole-trapping states into a BDP-OMe:$C_{60}$ bulk heterojunction via a low concentration (0.5 wt%) of m-MTDATA. While HBL-only reference devices display baseline PM (*EQE* ~350% at -4 V), adding m-MTDATA amplifies this threefold, achieving an *EQE* >1100%. At -2 V, *EQE* increases from 22% to 140%, yielding a specific detectivity of $4\times10^{12}$ Jones. Control experiments confirm that interfacial charge confinement remains a prerequisite for this gain. Importantly, all devices were fabricated exclusively by PVD, highlighting the compatibility of this approach with established vacuum-processing workflows and its relevance for scalable manufacturing and industrial upscaling.

Temperature-dependent admittance spectroscopy shows that the electrically active discrete immobilized-charge/trap-related response reaches its maximum at the optimized 0.5 wt% m-MTDATA loading. At higher concentration, this effective discrete response decreases, consistent with reduced electrical isolation of the m-MTDATA-related states and the associated suppression of PM gain. Complementary ultra-sensitive *EQE* measurements reveal pronounced sub-gap CT absorption in the 0.8–1.2 eV range, assigned primarily to m-MTDATA:$C_{60}$ CT states. This indicates that the performance decline at higher loading does not arise from the disappearance of CT states, but rather from a reduced fraction of states contributing to electrically effective charge retention needed to sustain interfacial space-charge accumulation. In parallel, the concentration-dependent increase in CT-related absorption correlates with the increase in dark current, consistent with field-assisted thermal charge generation through low-energy CT states. Despite this dark current rise, the gain enhancement at -2 V substantially exceeds the noise increase. With noise remaining below 50 fA $Hz^{-1/2}$, the device maintains a cutoff frequency $f_{-3dB}$ above 22 kHz. Taken together, these results establish a coherent mechanistic picture and identify precise control of trap concentration, such that electrically isolated trap states are maximized without entering a localized percolation regime, as a key design rule for next-generation trap-engineered PM-OPDs.

**4. Experimental Section**

*Device preparation:* As described before in Refs. [37,40,63] and reprinted here for completeness, a pre-structured ITO glass substrate was used to thermally evaporate all organic layers used in all of the devices in the vacuum chamber system (Kurt J. Lesker, UK) under ultra-high vacuum (pressure < $10^{-7}$ mbar). Prior to processing, substrates were thoroughly cleaned, and all organic materials underwent sublimation cleaning 1-2 times. An area of 6.44 $mm^2$ is defined by the overlap of the top and bottom electrodes. Devices were encapsulated with a glass cover and UV-hardened epoxy resin (Nagase ChemteX XNR 5592, Japan) after the evaporation process. To prevent the degradation of the devices, a moisture getter (Dynic Ltd. UK) was also used between the devices and the glass cover.

*Current-voltage measurements:* A source measurement unit (SMU) (Keithley SMU 2450, USA) was used to measure current-voltage characteristics under illumination and dark conditions. These devices were illuminated by a sun simulator (Solarlight Company Inc., USA) with an intensity of 100 mW $cm^{-2}$. At the same time, the light intensity was calibrated by a Hamamatsu S1337 silicon photodiode.

*Absorption spectra:* An UV-VIS-NIR spectrometer (SolidSpec-3700, Shimadzu Corporation, Japan) was used to characterize the optical properties of the organic thin films, including transmission (*T*) and reflection (*R*) by employing an integrating sphere. The absorption (*A*) spectrum, based on the measured transmission and reflection, was calculated with: $A = 1 - T - R$.

*External quantum efficiency:* A lock-in amplifier (Signal Recovery SR 7265) is used to measure the photocurrent generated by the device under monochromatic illumination from an Oriel Xe Arc-Lamp Apex Illuminator combined with a Cornerstone 260, 1/4 m monochromator (Newport, USA). The light is chopped at 170 Hz. The device signal is first amplified with a low-noise current preamplifier (SR570, Stanford Research Systems, USA) and converted into a voltage signal, which is then detected by the lock-in amplifier. The light intensity is monitored using the same measurement procedure and a calibrated silicon photodiode (Hamamatsu S1337 calibrated by Fraunhofer ISE). The *EQE* is determined from the ratio of the number of charge carriers generated by the device to the number of incident photons.

*Ultra-sensitive-External quantum efficiency: us-EQE* measurements were conducted under ambient conditions using a 250 W halogen lamp (OSRAM HLX 64657, Germany) as the excitation source. The light was chopped and coupled into a double monochromator (MSHD-300A, Quantum Design GmbH, Germany). The resulting monochromatic output was focused onto the OPD, and the generated photocurrent was recorded at -2 V. The current signal was first amplified with a current–voltage preamplifier (SR 570, Stanford Research Systems, USA) and subsequently analyzed using a lock-in amplifier (SR830, Stanford Research Systems, USA) with a time constant of 1 s. The incident photon flux was determined with calibrated Si and InGaAs photodiodes (Thorlabs FDS100-CAL, USA, and Hamamatsu G12183-020K, Japan), and the *EQE* was then obtained from the ratio of the OPD photocurrent to the measured photon flux. Measurement control was implemented via the SweepMe! software.

*Power spectral density:* An oscilloscope (DPO7354C, Tektronix USA) paired with a low-noise preamplifier (DLPCA-200, FEMTO Messtechnik GmbH, Germany) was used to obtain the noise spectra. The noise spectra are calculated from Welch's method.[64] To separate the actual noise data from the ambient environment, low-noise cables and shielding were installed. Both the number of data points and the sampling rate were tuned to guarantee a good resolution in the low-frequency range and a decent trend that extends to higher frequencies.

*Transient current measurements***:** The Transient rise and fall dynamics characteristics were measured using a 660 nm LED (M660L3, Thorlabs GmbH, Germany). An LED driver (DC2200, Thorlabs GmbH, Germany) controlled the modulated square wave light output and operated at 70 Hz. The photocurrent was then amplified (DHPCA-100, FEMTO Messtechnik GmbH, Germany) and recorded by the oscilloscope (Siglent SDS1204X-E, China).

**Supporting Information**

**Author Contribution**

A.S. and J.B. conceived and designed the project. A.S. designed the experiment, prepared the samples, conducted the measurements and data analysis of the PM-OPDs, and wrote the manuscript. K.L. and J.B. supervised the overall project. All authors contributed to reviewing and commenting on the data and the manuscript.

**Acknowledgments**

A.S. is financially supported by the doctoral scholarship from the Higher Education Commission of Pakistan (HEC), Pakistan, and the German Academic Exchange Service (DAAD), Germany. Additionally, the authors thank the Federal Ministry of Research, Technology, and Space (BMFTR) for the funding as part of the project to establish the German

Center for Astrophysics (03WSP1745). L.C.W. acknowledges the graduate academy project 2767 (GRK 2767), funded by the German Research Foundation (DFG). A.K. acknowledges support from the DFG project DANCE (LE 747/74-1). F.K. gratefully acknowledges support from the SAB project NDOT (Neuartige n-Dotanden für OPV und OLED, Project No. 100716679).

**Conflict of Interest**

The authors declare no conflicts of interest.

# Supporting Information

## Enhanced Photomultiplication Effect by Synergistic Integration of Hole-Blocking Layers and Trap Engineering in PM-OPDs

*Awais Sarwar[1,]*, Louis Conrad Winkler[1,2], Anncharlott Kusber[1], Fred Kretschmer[1], Karl Leo[1]*, Hans Kleemann[1], Johannes Benduhn[1,2,]**

### 1. Molecular structures

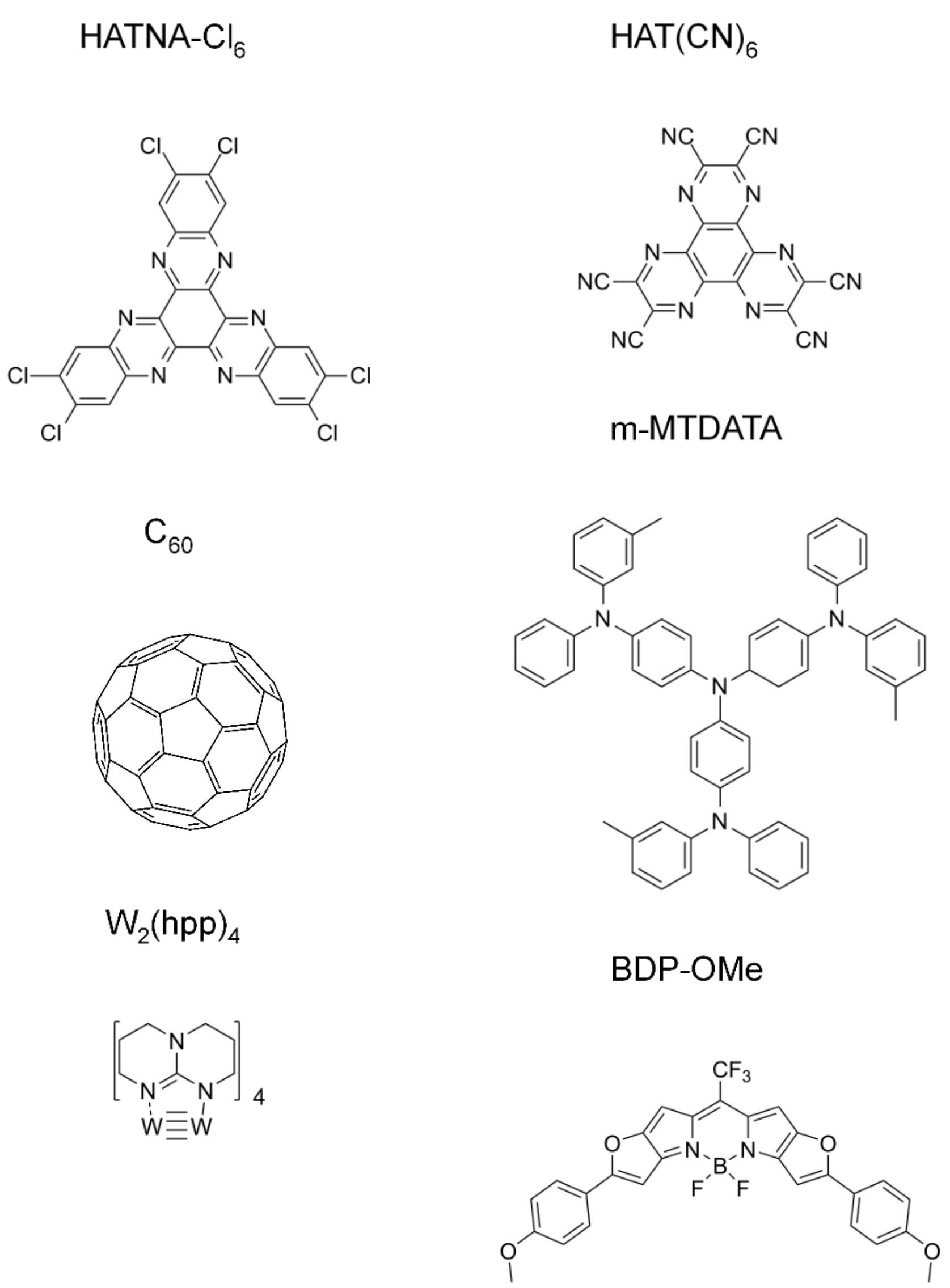

***Figure S1.*** *Molecular structures of materials used in this work.*

***Table S1.*** *List of materials used in this work.*

| **Material** | **Full name** | **Functionality** |
|---|---|---|
| $HAT(CN)_6$ | Hexaazatriphenylenehexacarbonitril | HBL |
| BDP-OMe | boron dipyrromethene with $CF_3$ on the meso-C | Donor |
| m-MTDATA | 4,4',4''-tris(3-methylphenylphenylamino)-triphenylamine | Hole trap |
| $C_{60}$ | Buckminsterfullerene | Acceptor/HBL |
| $W_2$(hpp)4 | Ditungsten tetra(hexahydropyrimidpyrimidine | n-dopant |
| HATNA-$Cl_6$ | 2,3,8,9,14,15-Hexachloro5,6,11,12,17,18-hexaazatrinaphthylene | ETL |

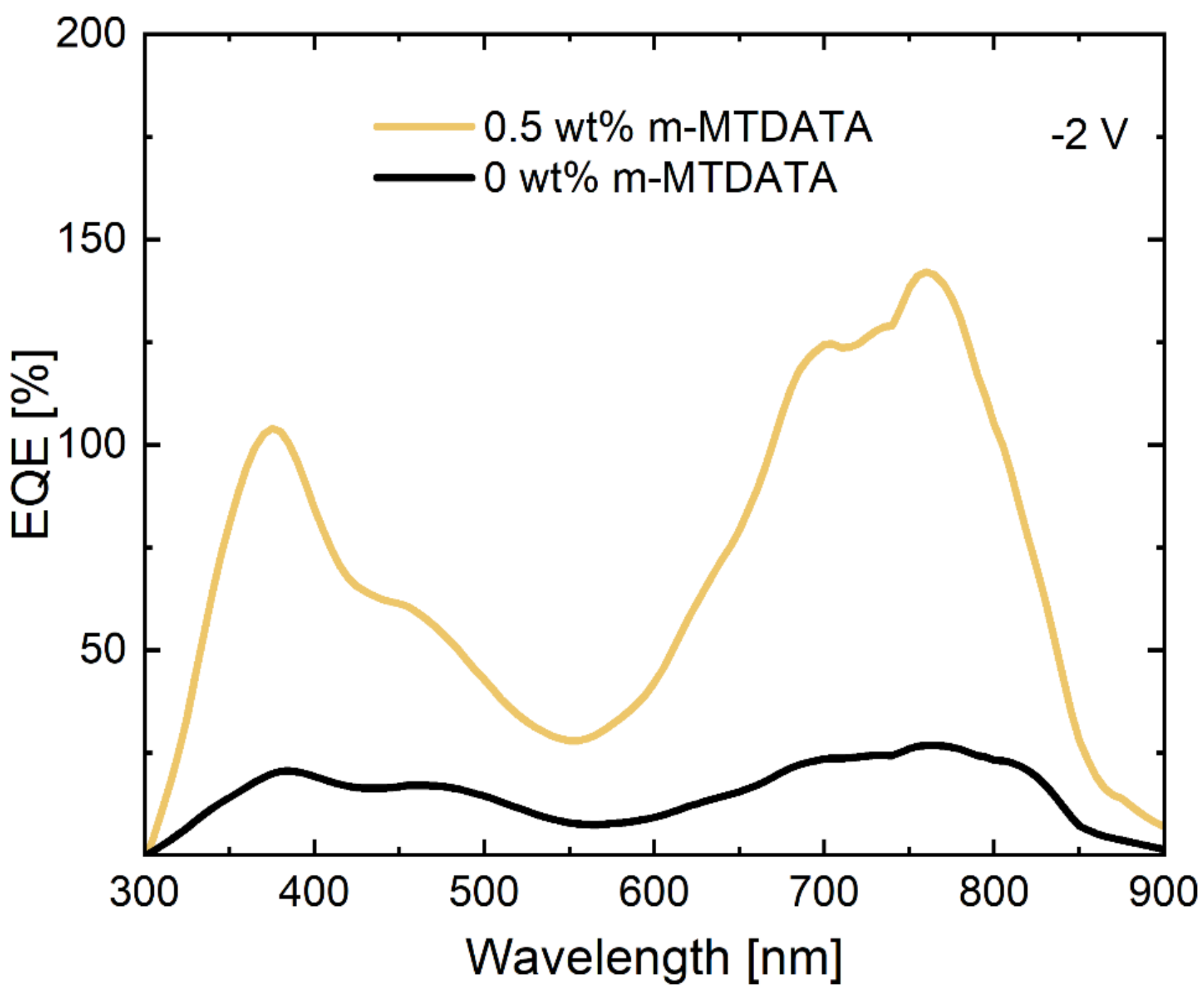

***Figure S2.*** *EQE spectra of devices incorporating hole-blocking layers (HBLs), comparing the reference device without m-MTDATA and the device containing 0.5 wt% m-MTDATA in the active layer, measured under -2 V reverse bias.*

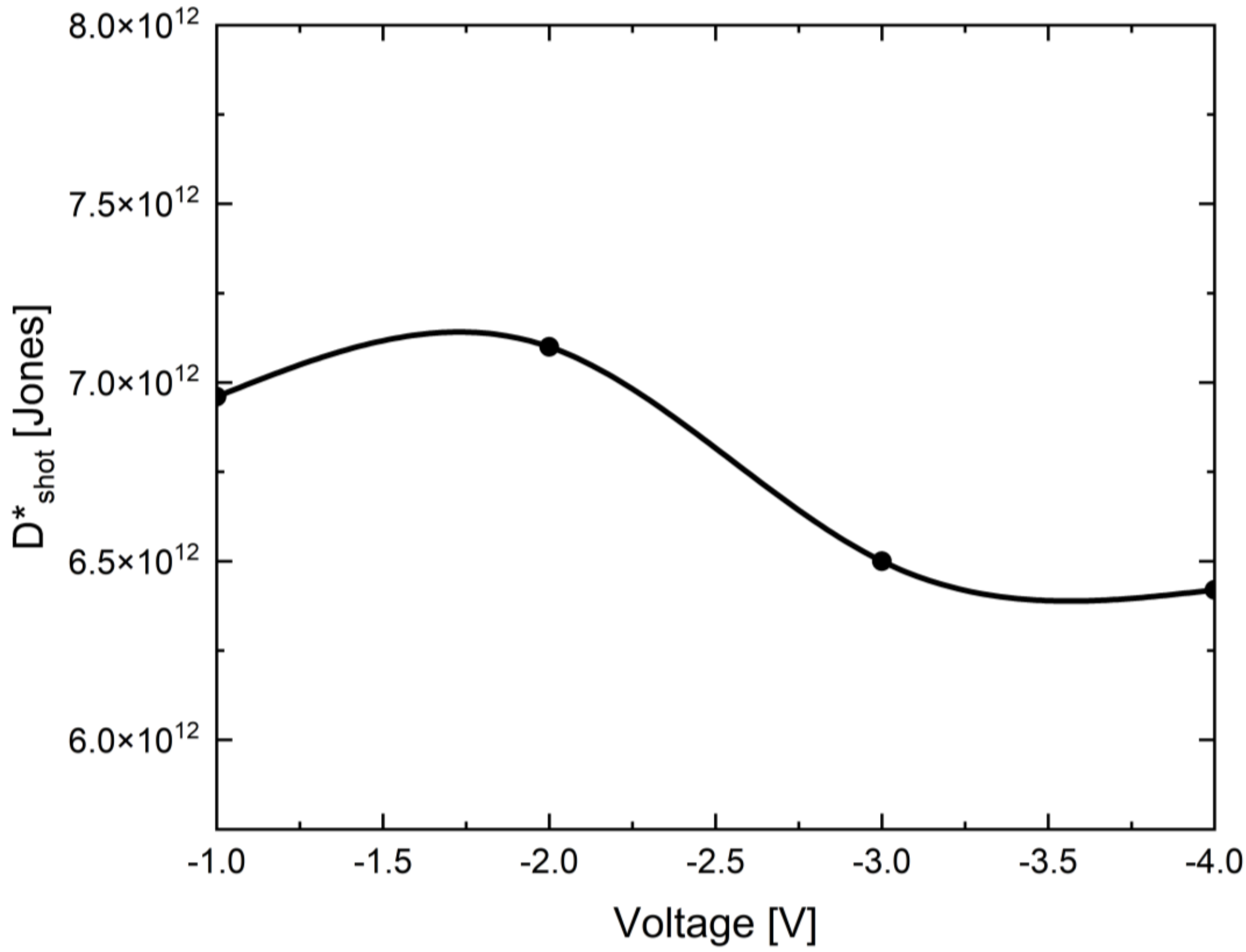

***Figure S3. Shot-noise-limited specific detectivity of the optimized PM-OPD.*** *Estimated shot-noise-limited specific detectivity, $D^{*}_{shot}$, of the 0.5 wt% m-MTDATA device as a function of reverse bias.*

***Table S2.*** *Extracted parameters from standard CT-state fits to the low-energy us-EQE response of devices with varying m-MTDATA concentration. The fitted quantities comprise the CT-state energy $E_{CT}$, reorganization energy $\lambda_{CT}$, and effective oscillator strength used to analyze the concentration-dependent absorption of the CT feature discussed in the main text.*

The low-energy us-EQE response was fitted using the Marcus-type CT-state absorption line shape:

$$\mathrm{EQE_{CT}}(E) = \frac{f_{\mathrm{CT}}}{E\sqrt{4\pi\lambda_{\mathrm{CT}}k_{\mathrm{B}}T}}\exp\left[-\frac{(E_{\mathrm{CT}}+\lambda_{\mathrm{CT}}-E)^2}{4\lambda_{\mathrm{CT}}k_{\mathrm{B}}T}\right]$$

Where $E_{CT}$, $\lambda_{CT}$ and $f_{CT}$ denote the CT-state energy, reorganization-energy parameter, and effective oscillator strength, respectively.[1–3]

| m-MTDATA concentration | $E_{CT}$ [eV] | $\lambda_{CT}$ [eV] | Effective oscillator strength [$eV^2$] |
|---|---|---|---|
| 0.1 wt% | 0.783 | 0.471 | 5.66e-06 |
| 0.2 wt% | 0.812 | 0.458 | 1.46e-05 |
| 0.5 wt% | 0.869 | 0.438 | 6.12e-05 |
| 1 wt% | 0.879 | 0.436 | 1.33e-04 |

**Estimation of m-MTDATA concentration from CT-fit oscillator strength**

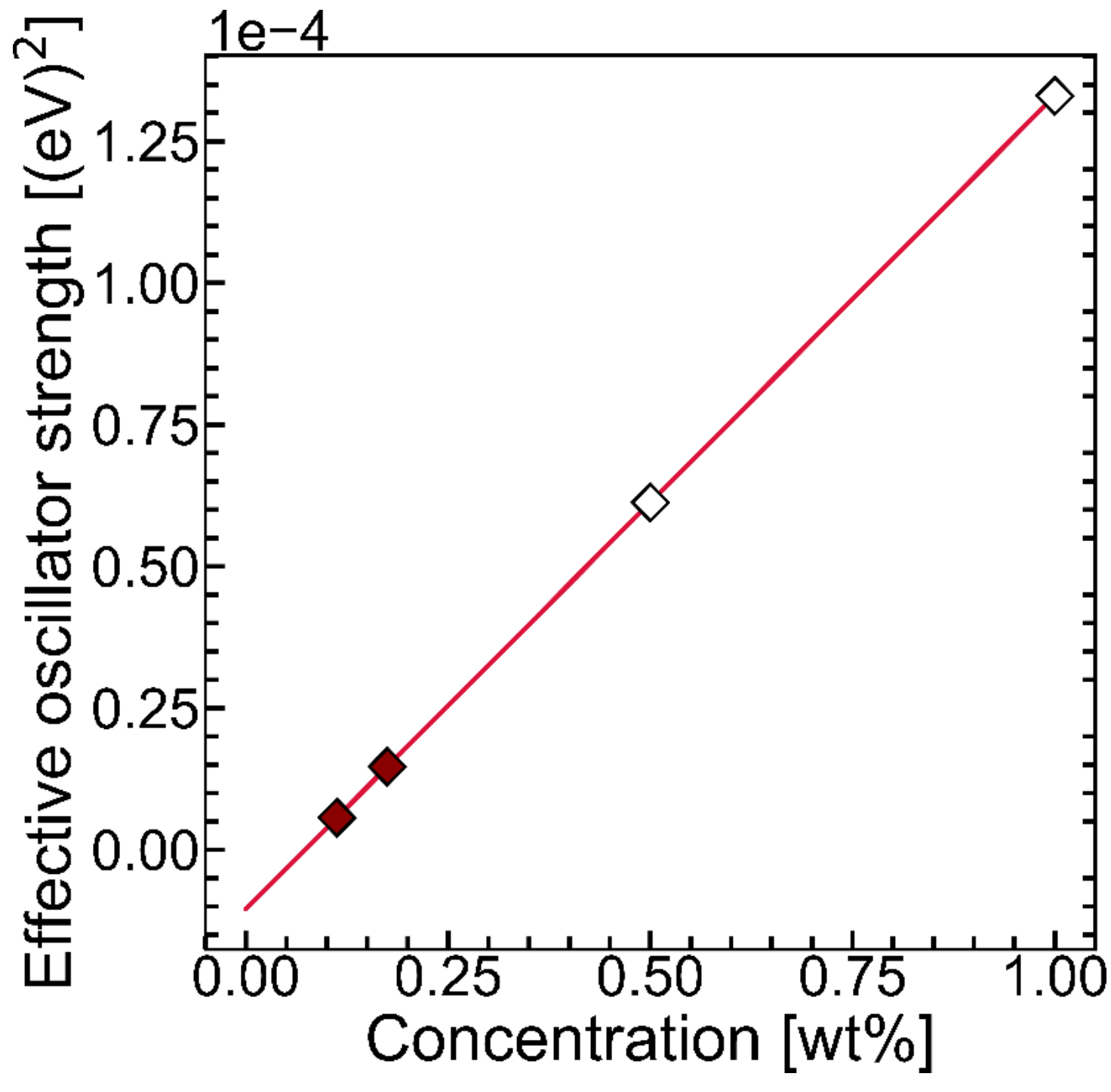

***Figure S4. Estimation of low m-MTDATA concentrations using CT-state oscillator strength.*** *The effective oscillator strength was determined by fitting the sub-gap feature with a Marcus line-shape function. The extracted values for samples containing m-MTDATA traps were then plotted as a function of trap concentration. Using BHJ samples with known m-MTDATA weight ratios (hollow symbols) as calibration references, the unknown low concentrations (filled symbols) were estimated from the linear relationship between effective oscillator strength and concentration in the low-concentration regime. This approach compensates for the limited process control associated with very low m-MTDATA stoichiometries during PVD co-evaporation.*

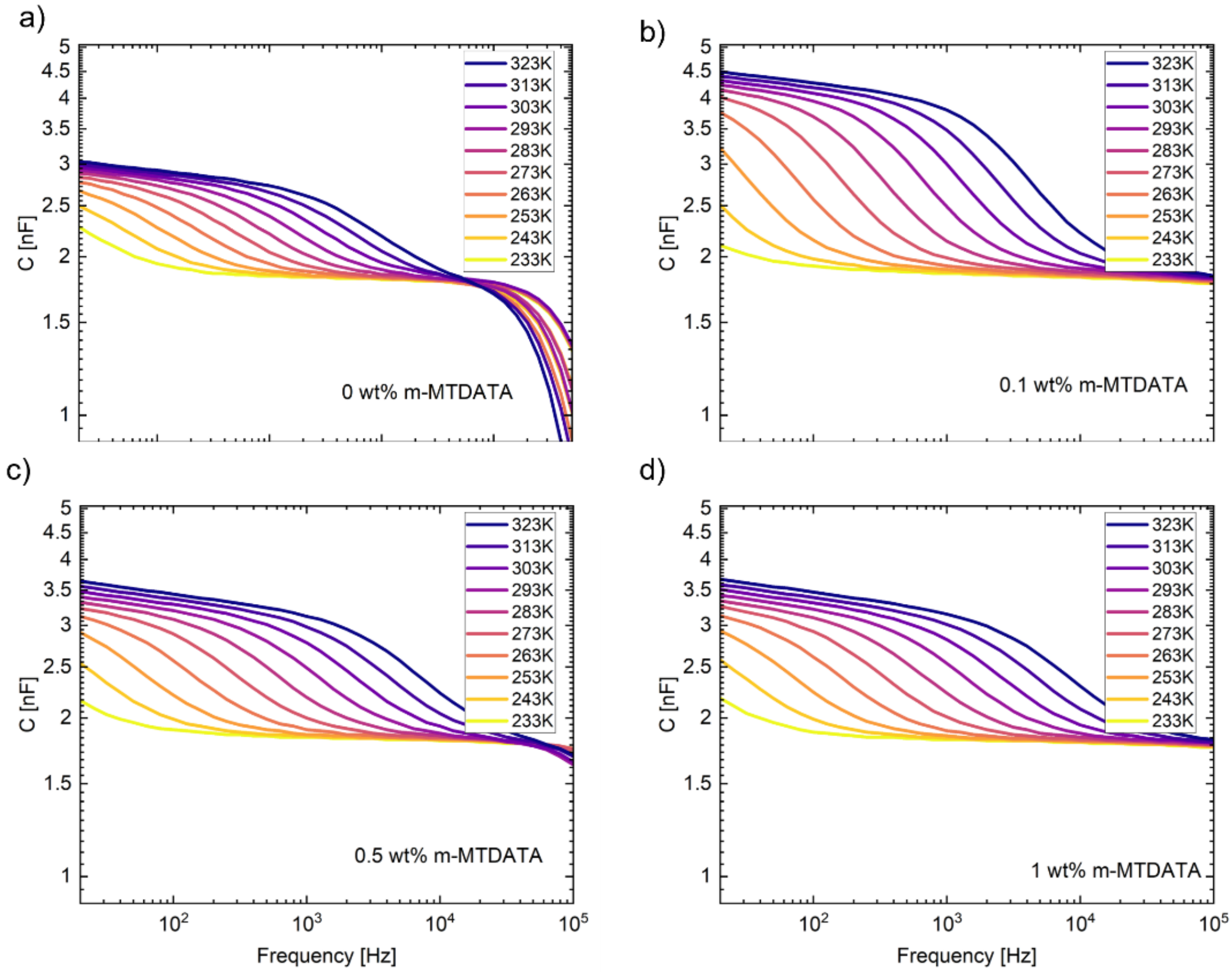

***Figure S5. Temperature-dependent capacitance–frequency characteristics used for admittance analysis.*** *Capacitance–frequency spectra of devices containing 0.0, 0.1, 0.5, and 1.0 wt% m-MTDATA, measured between 233 and 323 K. The observed temperature-dependent shift of the capacitance step to higher frequencies reflects a thermally activated charge response, commonly associated with electrically active trap-related states. This shift was used to estimate the corresponding activation energies and construct the admittance-derived profiles discussed in the main text.*

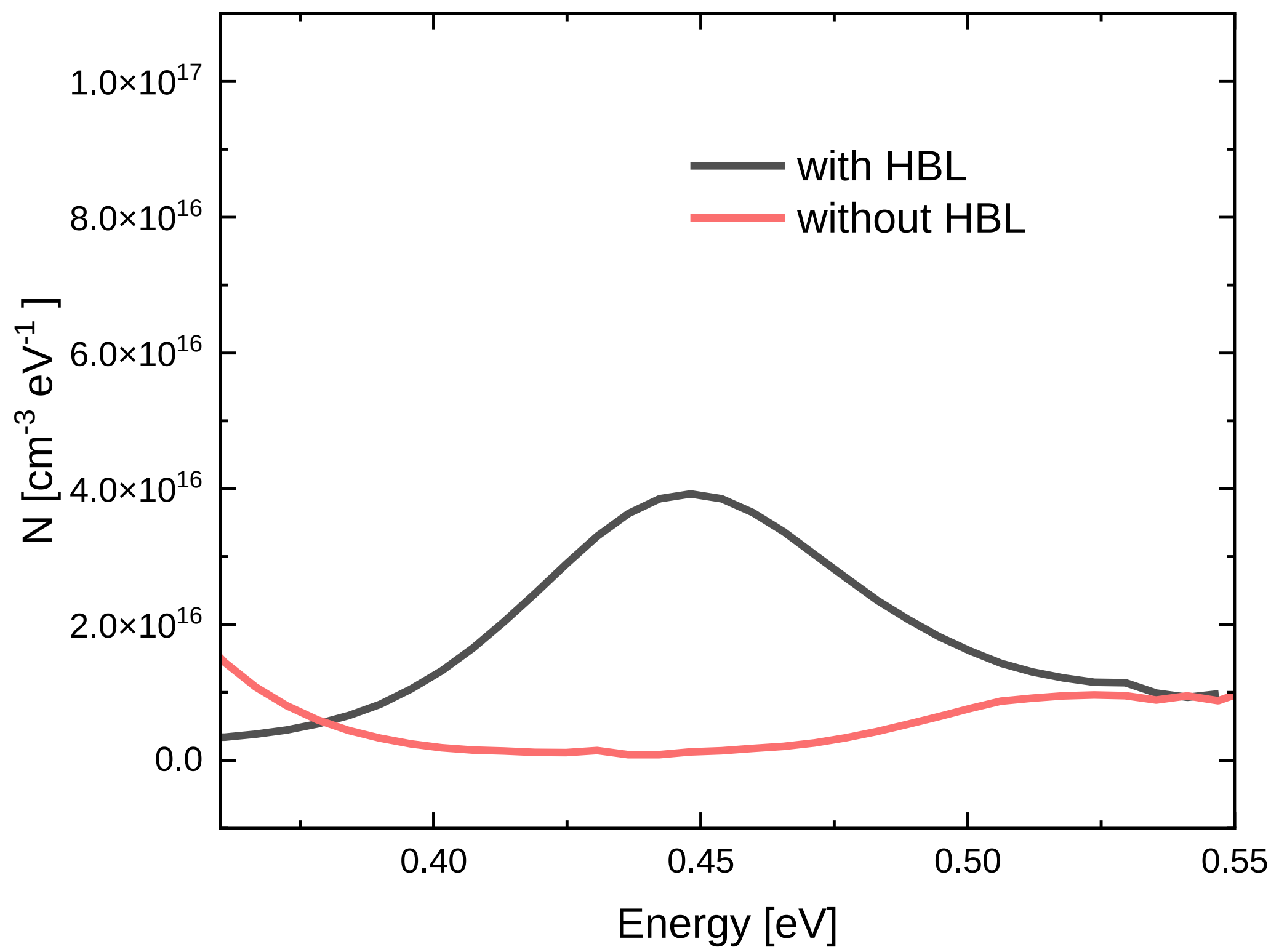

***Figure S6. Admittance-derived responses of devices without HBLs.*** *Immobilized-charge/trap-related profiles extracted from temperature-dependent capacitance–frequency measurements of no-HBL devices.*

**Supporting references***:*